\documentclass[prl,aps,twocolumn]{revtex4}
\usepackage{natbib}
\usepackage{amsmath} 
\usepackage{amssymb}	
\usepackage{graphicx} 
\usepackage{times}

\begin{document}
\newcommand\bbone{\ensuremath{\mathbbm{1}}}
\newcommand{\ul}{\underline}
\newcommand{\vl}{v_{_L}}
\newcommand{\vc}{\mathbf}
\newcommand{\be}{\begin{equation}}
\newcommand{\ee}{\end{equation}}
\newcommand{\bk}{{{\bf{k}}}}
\newcommand{\bK}{{{\bf{K}}}}
\newcommand{\cE}{{{\cal E}}}
\newcommand{\bQ}{{{\bf{Q}}}}
\newcommand{\br}{{{\bf{r}}}}
\newcommand{\bg}{{{\bf{g}}}}
\newcommand{\bG}{{{\bf{G}}}}
\newcommand{\hbr}{{\hat{\bf{r}}}}
\newcommand{\bR}{{{\bf{R}}}}
\newcommand{\bq}{{\bf{q}}}
\newcommand{\hx}{{\hat{x}}}
\newcommand{\hy}{{\hat{y}}}
\newcommand{\hd}{{\hat{\delta}}}
\newcommand{\bea}{\begin{eqnarray}}
\newcommand{\eea}{\end{eqnarray}}
\newcommand{\ra}{\rangle}
\newcommand{\la}{\langle}
\renewcommand{\tt}{{\tilde{t}}}
\newcommand{\upa}{\uparrow}
\newcommand{\dna}{\downarrow}
\newcommand{\bS}{{\bf S}}
\newcommand{\vS}{\vec{S}}
\newcommand{\dg}{{\dagger}}
\newcommand{\pdg}{{\phantom\dagger}}
\newcommand{\tphi}{{\tilde\phi}}
\newcommand{\cf}{{\cal F}}
\newcommand{\ca}{{\cal A}}
\renewcommand{\ni}{\noindent}
\newcommand{\ct}{{\cal T}}

\title{Spin-orbit coupled $j_{\rm eff}\!=\! 1/2$ iridium moments on the geometrically frustrated fcc lattice}
\author{A. M. Cook$^1$, S. Matern$^{2}$, C. Hickey$^1$, A. A. Aczel$^3$, and A. Paramekanti$^{1,4}$}
\affiliation{$^1$Department of Physics, University of Toronto, Toronto, Ontario, Canada M5S 1A7}
\affiliation{$^2$Institute for Theoretical Physics, Cologne University, 50937 Cologne, Germany}
\affiliation{$^3$ Quantum Condensed Matter Division,
Oak Ridge National Lab, Oak Ridge, TN, 37831, USA}
\affiliation{$^4$Canadian Institute for Advanced Research, Toronto, Ontario, M5G 1Z8, Canada}
\begin{abstract}
Motivated by experiments on the double perovskites La$_2$ZnIrO$_6$ 
and La$_2$MgIrO$_6$, we study the magnetism of spin-orbit coupled $j_{\rm eff} \! =\! 1/2$
iridium moments on the three-dimensional, geometrically frustrated, face-centered cubic lattice. The symmetry-allowed
nearest-neighbor interaction includes Heisenberg, Kitaev, and symmetric off-diagonal exchange.
A Luttinger-Tisza analysis shows
a rich variety of orders, including collinear A-type antiferromagnetism, stripe order with moments
along the $\{111\}$-direction, and incommensurate non-coplanar spirals, and we use Monte Carlo simulations
to determine their magnetic ordering temperatures. We argue that existing thermodynamic data on these iridates 
underscores the presence of a dominant Kitaev exchange, and also suggest a resolution to the puzzle
of why La$_2$ZnIrO$_6$, but not La$_2$MgIrO$_6$, exhibits `weak' ferromagnetism.
\end{abstract}
\maketitle

{\bf Introduction.} --- Heavy atoms with strong spin-orbit coupling (SOC) and electronic
correlations are predicted to form exotic quantum phases \cite{Krempa2014}.
Rare-earth ions with strong SOC on the frustrated
pyrochlore lattice can yield local moments with unusual exchange 
couplings, leading to `quantum
spin ice', as in Yb$_2$Ti$_2$O$_7$ \cite{Ross2011,SBLee2012,Shannon2012,Tchernyshyov2012,Armitage2014}.
Another exciting proposal is to realize the Kitaev Hamiltonian, with a spin liquid ground state and Majorana
fermion excitations \cite{Kitaev2006},
in iridium oxides with edge-sharing octahedra, such as 
the two-dimensional (2D) honeycomb iridates Na$_2$IrO$_3$ and Li$_2$IrO$_3$ \cite{Jackeli2009,Jackeli2010}.
Doping such Mott insulators has been predicted to lead to topological
superconductivity \cite{You2012,Hyart2012,Okamoto2013a,Okamoto2013b,Scherer2014}.
Experimentally, in both Na$_2$IrO$_3$ and Li$_2$IrO$_3$, the spin liquid state is preempted by
magnetic order \cite{YJKim2011,Coldea2012} induced by interactions beyond the Kitaev model. 
Nevertheless, extensive work on these materials
\cite{YSingh2012,Brink2014,Rau2014a,Rau2014b,Rachel2014,Perkins2014}, and
3D harmonic honeycomb iridates $\beta,\gamma$-Li$_2$IrO$_3$ 
\cite{Takagi2014,Analytis2014,Coldea2014a,Kimchi2014b,YBKim2014a,YBKim2014b,YBKim2014c,YBKim2015a},
ascribes their complex order to large Kitaev couplings.
Kitaev interactions in the triangular iridate Ba$_3$IrTi$_2$O$_9$
may lead to vortex crystals or gauge-like degeneracies \cite{Daghofer2012,Trebst2014,Jackeli2015}.

In light of these studies, we explore the following important issues. 
What kinds of phases does the Kitaev interaction support in 3D lattices with geometric frustration? Do
experiments suggest dominant Kitaev interactions in any geometrically frustrated materials?
Here, we address these questions in the context of ordered double perovskite (DP) compounds,
a large class of materials with the chemical formula A$_2$BB'O$_6$, 
where B and B' ions occupy the two sublattices of a 3D cubic crystal.  
Metallic DPs such as
Sr$_2$FeMoO$_6$ \cite{kobayashi1998} are of great interest as half-metallic ferromagnets 
\cite{sarma2000,sahadasgupta2001,brey2006,alff2007,erten2011}.
Recent work on metallic DPs has examined the role of SOC on bulk spin dynamics \cite{Plumb2013},
and Chern bands in ultrathin films \cite{Cook2013,Liang2013,Cook2014a,Cook2014b,Vanderbilt2014}.
On the other hand, DPs where
B is an inert filled-shell ion, and B' is a heavy $4d$/$5d$ ion, form Mott insulators with local moments
on the frustrated fcc lattice of B' ions \cite{aharen2010,aharen2010n2,devries2010,steele2011,aczel2013,Chen2010,Dodds2011,Chen2011,Ishizuka2014}.
Our work is motivated by the recent synthesis of La$_2$ZnIrO$_6$ and La$_2$MgIrO$_6$ \cite{cao2013}.
Structurally, both materials have nearly undistorted oxygen octahedra. A nominal valence Ir$^{4+}$
($5d^5$), together with the strong SOC and larger spacing between Ir ions compared to perovskites, 
suggests that these materials behave as effective $j_{\rm eff}\!\!=\!\!1/2$ Mott insulators \cite{cao2013}.

In this Rapid Communication, we focus on the broad aspects of magnetism in an ideal fcc lattice,
highlighting the rich physics of strong SOC in a canonical frustrated 3D lattice.
Our key results are
the following. (i) We show that even the nearest-neighbor symmetry-allowed
Hamiltonian on the fcc lattice,
which includes Heisenberg, Kitaev, and symmetric off-diagonal exchange couplings, leads to rich magnetic phases
such as collinear antiferromagnetism, stripes, or multimode spirals.
 Indeed, previous work \cite{Kimchi2014a} has suggested that strong Kitaev interactions 
should be present in a large class of 2D and 3D lattices, including the fcc lattice, but did not study the most
general symmetry-allowed Hamiltonian. (ii) We find that strong SOC can also stabilize a regime of robust 
A-type antiferromagnetism (AFM), also called Type-I AFM, which is observed in
neutron diffraction on La$_2$ZnIrO$_6$ and La$_2$MgIrO$_6$ \cite{cao2013}.
Our results challenge the conventional wisdom which ascribes robust A-type antiferromagnetism in many fcc magnets to 
further neighbor Heisenberg exchange \cite{98_seehra,01_lefmann}, and suggests that anisotropy due to SOC may be crucial in $5d$ oxides.
Indeed, a recent {\it ab initio} study of Sr$_2$CrSbO$_6$ \cite{tanusri2012} finds next-neighbor interactions are negligible,
$\lesssim 5\%$ of the first neighbor interactions. (iii) In certain regimes
with $A$-type AFM, we uncover a  residual accidental XY degeneracy of collinear states. Thermal order by disorder pins the
moments along the Ir-O bond directions.
(iv) We argue that thermodynamic data on La$_2$ZnIrO$_6$ and La$_2$MgIrO$_6$ \cite{cao2013}, i.e., their ordering pattern and 
small frustration parameter, indicate a dominant antiferromagnetic Kitaev coupling. Microscopically, 
this may arise from the near-cancellation of Heisenberg interactions, from multiple Ir-O-O-Ir 
superexchange paths
\cite{Jackeli2009,Jackeli2010,Kimchi2014a,Trebst2014}, and the smaller direct exchange for well-separated Ir 
atoms in the DP structure. We argue that a
subtle difference in magnetic orders can reconcile
`weak' ferromagnetism in La$_2$ZnIrO$_6$ with its absence in La$_2$MgIrO$_6$.  These
compounds thus realize a new class of `Kitaev materials'. Ultrathin films of La$_2$BIrO$_6$, grown along \{111\}, 
could realize the triangular lattice AFM Kitaev model.

{\bf Model}. --- To construct a minimal model on the fcc lattice of Ir moments, we consider the ideal cubic DP structure, and
focus on nearest neighbor terms which are expected to dominate. We appeal to symmetry arguments
to write down all possible terms, based on the fact that the effective $j_{\rm eff}\!=\!1/2$ angular momentum operator
is a pseudovector (axial vector). Requiring invariance of the Hamiltonian under lattice rotational and mirror symmetries
\cite{Halg1986}
constrains the Hamiltonian coupling nearest-neighbor Ir sites to be of the form $H=H_{\rm H}+H_{\rm K} +H_{\rm OD}$,
\bea
\!\! H_{\rm H} \!&=&\! J_H \!\! \sum_{\la\br \br'\ra} \vec S_\br \cdot \vec S_{\br'} \\
\!\! H_{\rm K} &=& J_K (
\!\!\!  \sum_{\la \br\br'\ra_{xy}}\!\!\!\! S^z_\br S^z_{\br'}+
\!\!\!  \sum_{\la \br\br'\ra_{yz}} \!\!\!\!  S^x_\br S^x_{\br'}+
\!\!\!  \sum_{\la \br\br'\ra_{xz}} \!\!\!\!  S^y_\br S^y_{\br'}) \\
\!\! H_{\rm OD} &=& \Gamma \sum_\br
\left[ (S^x_{\br} S^y_{\br+x+y}  + S^y_{\br} S^x_{\br+x+y} - 
S^x_{\br} S^y_{\br-x+y} \right. \nonumber \\
&-& \left. S^y_{\br} S^x_{\br-x+y}) + (x,y \leftrightarrow y,z) + (x,y \leftrightarrow x,z)  \right].
\eea
Here, $\la \br \br'\ra$ denotes all first-neighbor pairs, while $\la \br \br'\ra_{xy}$ denotes first-neighbors restricted to the $xy$-plane
(similarly for $yz,xz$). $H_{\rm H}$ is the Heisenberg term, $H_{\rm K}$ is the Kitaev interaction, and  $H_{\rm OD}$ is a 
symmetric off-diagonal exchange term. Antisymmetric Dzyaloshinskii-Moriya interactions are forbidden here by inversion symmetry.
A dominant $J_H < 0$ leads to ferromagnetism; this is incompatible with the ordering observed
in La$_2$BIrO$_6$ (B=Mg,Zn), so we assume $J_H > 0$.

\begin{figure}[t]
\includegraphics[scale=0.7]{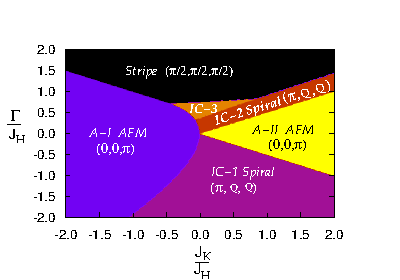}
\caption{Phase diagram of the nearest-neighbor spin Hamiltonian as a function of $J_K/J_H$ and $\Gamma/J_H$, obtained
using the Luttinger-Tisza (LT) method. The AFM states are A-type antiferromagnets, having ferromagnetic planes stacked
antiferromagnetically along the third direction, with spins either pointing perpendicular to the FM plane (A-I AFM) or lying 
in the FM plane (A-II AFM). Stripe order at $(\pi/2,\pi/2,\pi/2)$ features
moments pointing in the $\{111\}$ or symmetry related directions. IC-1, IC-2, and IC-3  are 
incommensurate noncoplanar spirals; beyond the LT analysis, they are multimode states.}
\label{fig:LT}
\end{figure}

{\bf Luttinger-Tisza analysis.} ---
To determine the preferred magnetic orders, we use the Luttinger-Tisza (LT) method which considers the spins to be
classical moments, and replaces the
constant length spin vectors by unconstrained vector fields $\vec\phi_\br$. 
The classical spin Hamiltonian written in momentum space
then takes the form
$
H_{\rm LT} = 2 J_H \sum_\bk \phi_{\bk\mu}^* M^\pdg_{\mu\nu}(\bk) \phi^\pdg_{\bk\nu}
$
with
\bea
M(\bk) = \begin{pmatrix} A_\bk + \alpha C^{yz}_{\bk} &- \gamma S^{xy}_{\bk} & - \gamma S^{xz}_{\bk} \\ - \gamma S^{xy}_{\bk} & A_\bk + \alpha C^{xz}_{\bk} & - \gamma S^{yz}_{\bk}\\
- \gamma S^{xz}_{\bk} & - \gamma S^{yz}_{\bk} & A_{\bk} + \alpha C^{xy}_{\bk} \end{pmatrix}.
\eea
Here,
$A_\bk = (\cos k_x \cos k_y + \cos k_x \cos k_z + \cos k_y \cos k_z)$, 
$C^{ij}_{\bk} = \cos k_i \cos k_j$, and $S^{ij}_{\bk} = \sin k_i \sin k_j$, and we have defined $\alpha=J_K/J_H$ and $\gamma=\Gamma/J_H$.
Here, $k_i$ (with $i=x,y,z$) denote components of the momentum along the cubic Ir-O axes, and we have set the Ir-O-B bond length (B=Zn,Mg)
to unity.
Diagonalizing $H_{\rm LT}$ for $J_H > 0$, and looking for the lowest energy eigenvalue in $\bk$, we find the rich variety of magnetic
orders shown 
in Fig.~\ref{fig:LT}.

{\bf Magnetic orders.} --- The LT analysis yields collinear as well as spiral antiferromagnetic (AFM) states.
We describe these phases below, and compare their energy with numerical simulated annealing results.

\begin{figure}[t]
\includegraphics[scale=0.22]{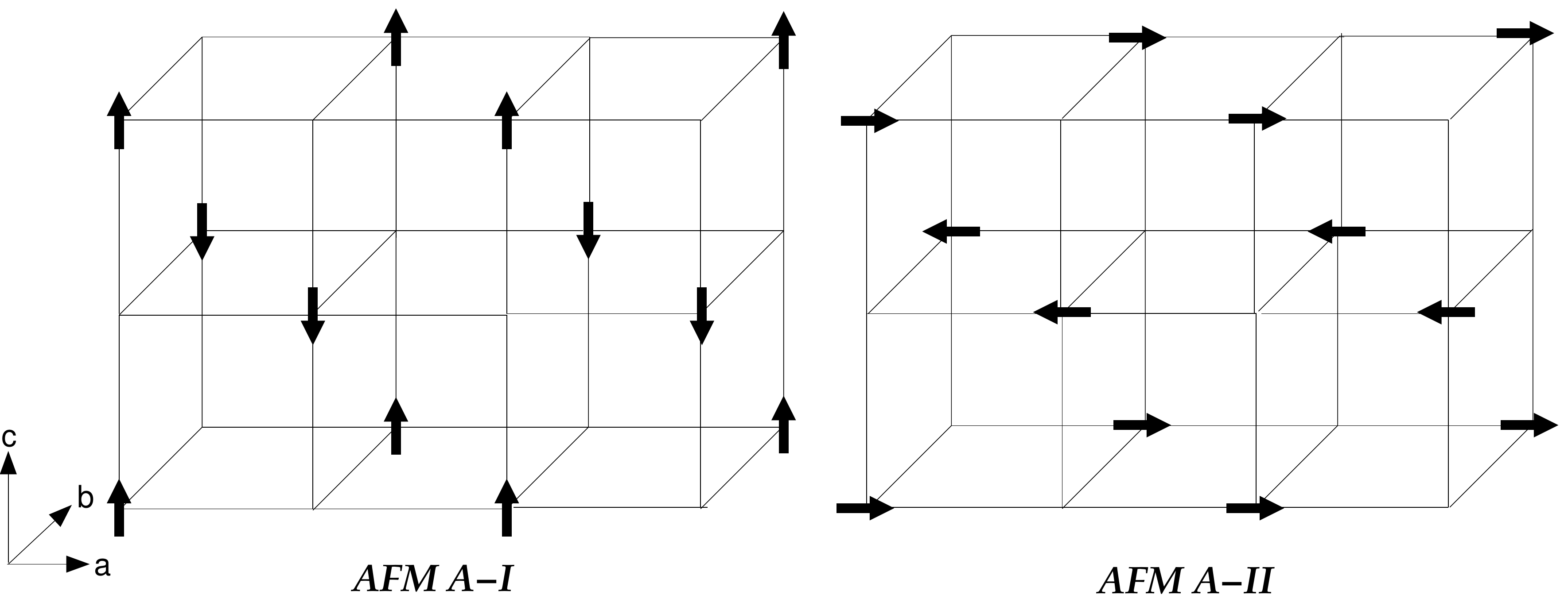}
\caption{Real space spin configurations in the layered A-type antiferromagnetic states AFM A-I and AFM A-II.}
\label{fig:afm}
\end{figure}

{\it A-I AFM:} This is an A-type collinear AFM (also referred to as a Type-I AFM in the literature) which consists of 
ferromagnetically ordered spins in the cubic ab-plane 
layered antiferromagnetically along the  c-axis . The spins point along the c-axis, perpendicular to the ferromagnetic planes
as shown in Fig.~\ref{fig:afm}.
There are six symmetry related A-I AFM ground states, associated with a three-fold choice of the layering direction and a two-fold
choice of the Ising AFM order. Although these are the lowest energy collinear states, there is an accidental classical
degeneracy, where one can form multimode states leading to coplanar or even noncoplanar states
with the same classical ground state energy.
This degeneracy is expected to be broken in favor of collinear states by fluctuation effects, and our simulated annealing finds 
the above collinear states to be stabilized by thermal order by disorder.

{\it A-II AFM:} This is also an A-type collinear antiferromagnet; however spins lie in the ferromagnetic planes as in Fig.~\ref{fig:afm}. 
In addition to collinear states, there are again multimode
coplanar or noncoplanar states with the same classical ground state energy; we expect and observe numerically that thermal 
fluctuations favor the collinear orders. However, the ground
state energy is independent of the precise angle in the plane so that there is an accidental XY degeneracy of collinear states. 
Our simulated annealing results show that this degeneracy is also broken by thermal fluctuations, with `order by disorder' favoring 
spins along the Ir-O bond direction. There are twelve symmetry related 
A-II ground states favored by fluctuations, arising from a three-fold choice of the layering direction and a four-fold choice of the spin axis.
{\it Remarkably, the A-II AFM order persists even in the pure Kitaev limit with} $J_K > 0$.

\begin{figure}[t]
\includegraphics[scale=0.22]{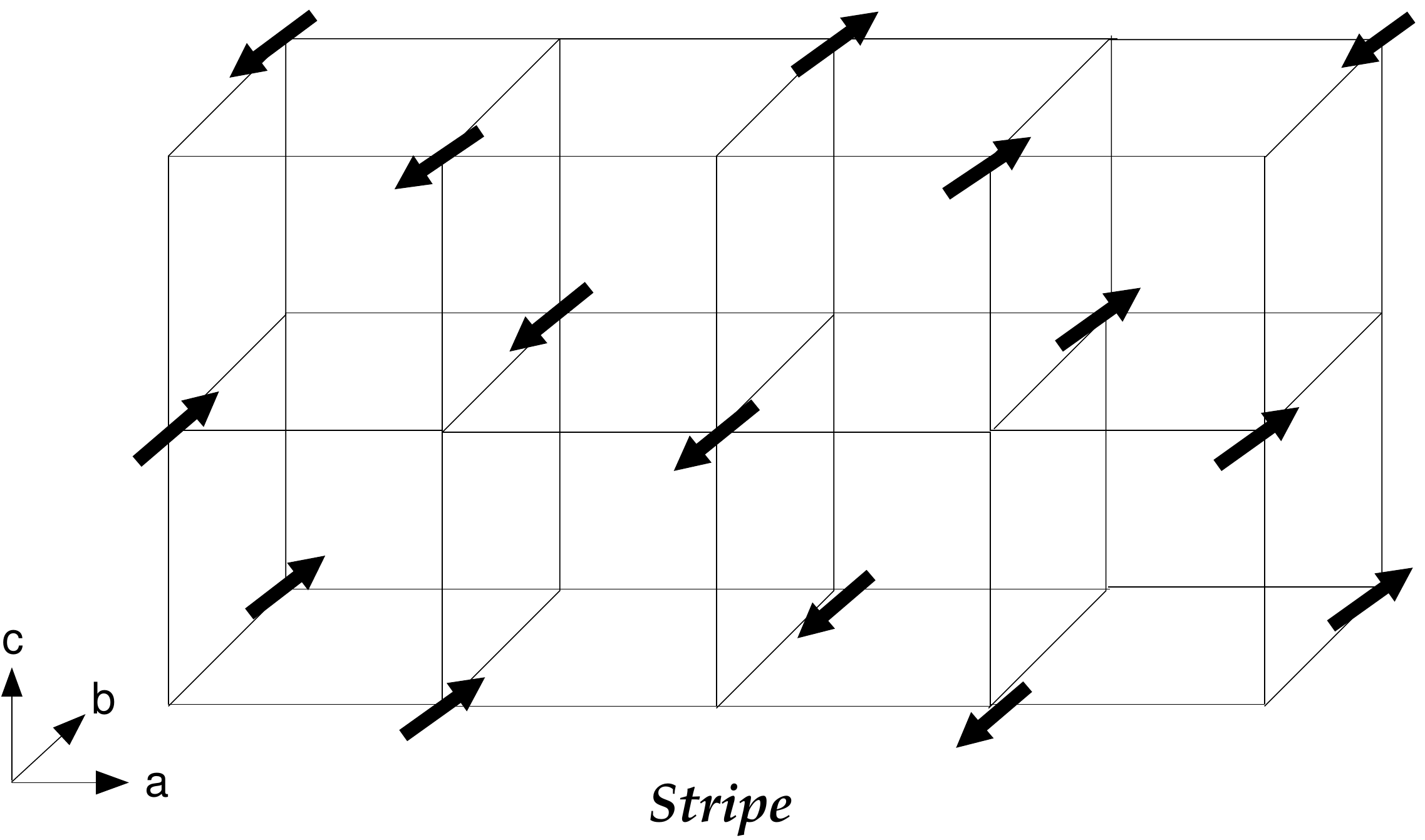}
\caption{Real space spin configurations in the collinear stripe state, showing moments pointing along the
diagonal $\{111\}$ and $\{\bar{1}\bar{1}\bar{1}\}$ directions for $(k_x,k_y,k_z) \equiv (\pi/2,\pi/2,\pi/2)$.}
\label{fig:stripe}
\end{figure}

{\it Stripe:} The collinear stripe state has spins pointing along the $\{111\}$ and $\{\bar{1}\bar{1}\bar{1}\}$ directions
arranged as shown in Fig.~\ref{fig:stripe} for $(k_x,k_y,k_z) \! \equiv \! \pm (\pi/2,\pi/2,\pi/2)$; symmetry related orders are degenerate.
Ordering with this wavevector is also referred to as a Type-II AFM.
The ordering wavevector determines the direction of the spins, so that flipping one of the momentum components also flips
the corresponding spin component; ordering at $\pm (\pi/2,-\pi/2,\pi/2)$ leads to spins along $\{1\bar{1}1\}$ and 
$\{\bar{1} 1 \bar{1}\}$. This leads to a total of eight ground states.

{\it Incommensurate Spiral (IC-1, IC-2):} In these regimes, the LT analysis suggests an incommensurate coplanar 
spiral order with wavevector $(k_x,k_y,k_z) \equiv (\pi,Q,Q)$, and symmetry related equivalents. 
With $\alpha=J_K/J_H$ and $\gamma=\Gamma/J_H$,
minimizing
the LT energy leads to $Q \!=\! \cos^{-1}(\frac{1+\alpha/2}{1+|\gamma|})$; the transition into the AFM A-I state
($Q\!=\! 0$) happens when $\alpha\!=\! 2|\gamma|$. However, we find that if we assume single mode ordering, the spins
constructed in the IC-1 and IC-2 phases from the LT eigenvectors do not satisfy the constraint of constant 
magnitude. Our simulated annealing numerics show that the ground states in this 
regime are noncoplanar multimode spirals formed by superposing all six equivalent wavevectors $(\pi,Q,\pm Q), (Q,\pi,\pm Q), (Q,\pm Q,\pi)$.

{\it Incommensurate Spiral (IC-3):} In this regime,  the LT approach again suggests an incommensurate coplanar 
spiral order; however, the wavevector is of the form $(k_x,k_y,k_z) \equiv (P,Q,Q)$. We have not found a simple closed
form expression for $P,Q$; however, they are obtained by minimizing the LT eigenvalue
\bea
\!\!\!\!\! \lambda \!\! &=& \!\! (4 \!+\! \alpha) \cos P \! \cos Q \!+\! (2 \!+\! \alpha) \cos^2 \!Q  \!-\! \gamma \sin^2\! Q \!-\! \sqrt{D} \\
\!\!\!\!\! D \!\!\! &\equiv & \!\!\! [\alpha (\cos \! P \!-\! \cos \! Q) \! \cos \! Q  \!-\! \gamma \sin^2 \! Q]^2\! \!+\! 8 \gamma^2 \! \sin^2\!P \! \sin^2\!Q
\eea
Again, a single mode spiral does not satisfy the spin constraint, and our simulated annealing numerics show
noncoplanar multimode spiral order in this regime.

\begin{figure}[t]
\includegraphics[scale=0.35]{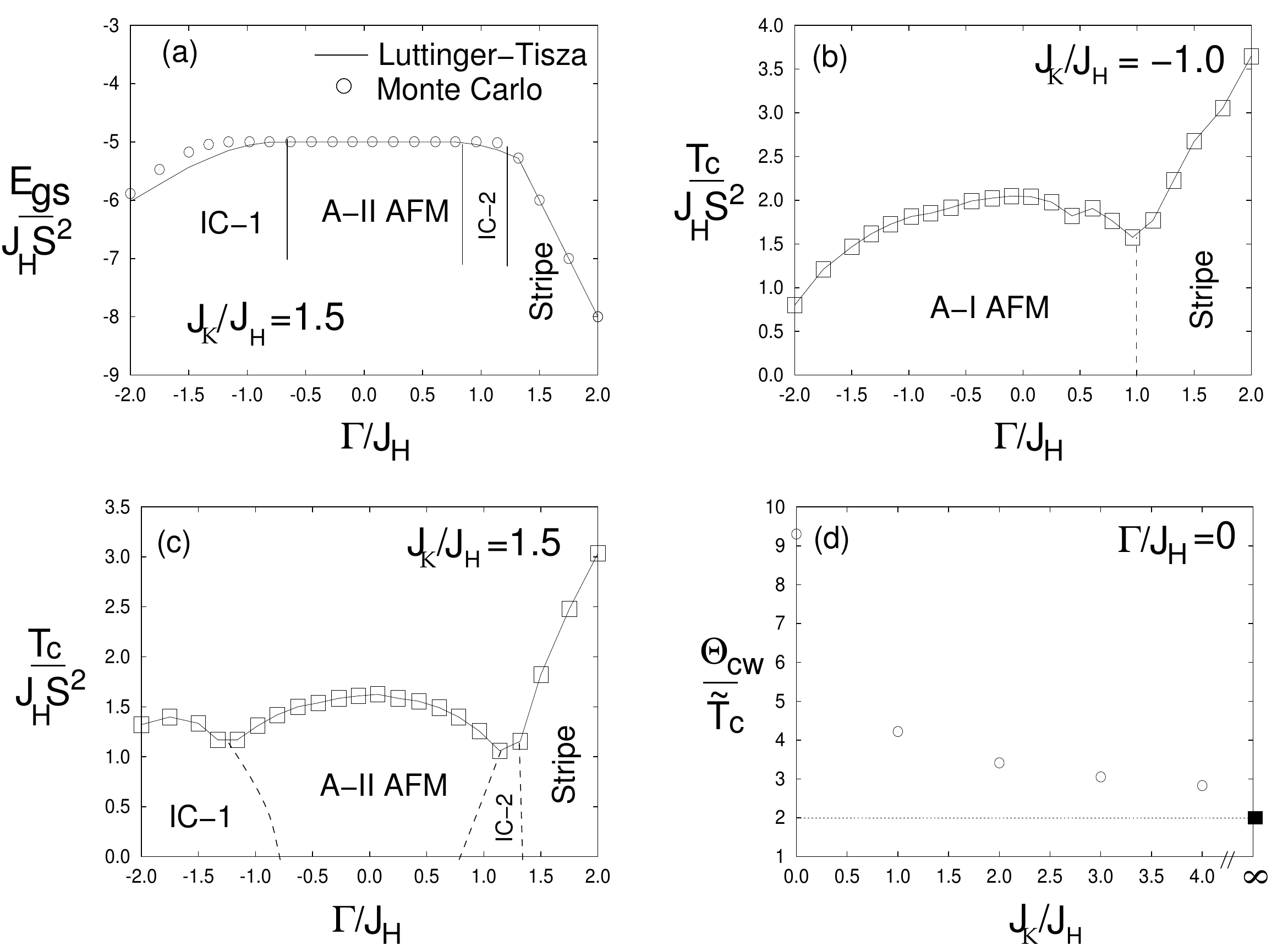}
\caption{(a): Comparison of the ground state energy per spin $E_{gs}$ obtained within LT method (solid line)
and simulated annealing (dots). (b),(c): Magnetic transition temperature $T_c$ of the classical model (in units of $J_H 
S^2$, for spin length $S$) vs. $\Gamma/J_H$, obtained using Monte Carlo simulations for cuts through the 
phase diagram (Fig.~\ref{fig:LT}) at $J_K/J_H \!=\! -1.0,+1.5$. (d) Plot of the ``frustration parameter'', the ratio of the 
$\tilde{T}_c \equiv T_c (1+1/S)$ to $\Theta_{CW}$; the rescaling of $T_c$ by $(1+1/S)$ accounts for the classical 
$S^2$ being replaced by the quantum $S(S+1)$. The dark square shows the result at $J_K/J_H \to \infty$.}
\label{fig:Tc}
\end{figure}

{\bf Monte Carlo results.} --- To complement the LT analysis, we have used simulated annealing numerics,
which preserves the spin constraint, to
find the classical ground states. Fig.~\ref{fig:Tc}(a) compares the numerically computed ground state
energy per spin to the Luttinger-Tisza result, for $J_K/J_H=1.5$ and varying $\Gamma/J_H$. The agreement
between the two is excellent in the A-II AFM and Stripe states, where the collinear order is precisely recovered. 
Our result that the A-II AFM state appears even for large $J_K$ differs from an earlier study \cite{Kimchi2014a}
which proposed a spiral ground state based on a Luttinger-Tisza analysis which did not take into account
thermal fluctuations and order-by-disorder.
For IC-1/IC-2, the simulations indicate multimode order, and lead to an energy per spin  (for $36^3$
lattice) which is only slightly higher by $\lesssim 2\%$.

In order to determine the magnetic ordering temperature in the various phases, we
have carried out Monte Carlo simulations on system sizes with up to $24^3$ spins. Fig.~\ref{fig:Tc} shows 
the magnetic $T_c$ as determined from the specific heat singularity, along various cuts through the Luttinger-Tisza 
phase diagram. The Heisenberg limit in
the absence of SOC ($J_K \! = \!0$, $\Gamma  \! =  \! 0$) is the most fragile state with the lowest $T_c
\approx 0.44 J_H S^2$; our results here agree with previous work on the fcc Heisenberg model \cite{Diep1989},
where thermal order by disorder leads to a nonzero $T_c$.
The A-I AFM, A-II AFM, and stripe phases appear most robust with high $T_c$, since
SOC enhances the pinning of the moment direction.  Thus,
although the exchange interactions induced by SOC are frustrated on the fcc lattice, the SOC nevertheless 
enhances $T_c$ by favoring certain spin orientations, thus reducing the effects of thermal disordering.

{\bf Comparison with experiments.} --- La$_2$ZnIrO$_6$ and La$_2$MgIrO$_6$ are A-type AFMs.
Combined {\it ab initio} and neutron diffraction
studies \cite{cao2013} suggest that the Ir spins lie predominantly in the ferromagnetic planes, viz. the A-II AFM state.
This is consistent with $J_K > 0$ and $|\Gamma| < J_K /2$. Order by disorder
pins moments along the Ir-O bond directions.

The Curie-Weiss 
temperature of $j_{\rm eff}\!=\!1/2$ moments on the ideal fcc lattice is $\Theta_{CW} \!  = \!  - (3 J_H \!  + \!  J_K)$, independent of $\Gamma$.
However, both La$_2$ZnIrO$_6$ and La$_2$MgIrO$_6$ have a monoclinic P$2_1$/n structure, arising from
small IrO$_6$ octahedral rotations --- an octahedral rotation $\phi$ about the cubic $c$-axis which is staggered
between adjacent $ab$ layers, and a global tilt about the cubic $\{110\}$ axis. In the strong SOC limit,
the Ir moments track the
octahedral rotation, as shown for Sr$_2$IrO$_4$ \cite{Jackeli2009,Senthil2011}. 
A high temperature expansion yields a powder averaged $\Theta_{CW} \! = \! - J_H  \!  - \!  \frac{1}{3} (2 J_H \! + \! J_K)  (1 \!  + \!  2\cos 2\phi)$.
If the axis along which the ferromagnetic planes are stacked in staggered fashion
coincides with the axis of the staggered octahedral rotations, it leads to a net ferromagnetic 
moment $\approx m \sin\phi$ in the A-II AFM state, where $m$ is the ordered moment. Equivalently,
we may start with the Hamiltonian in the ideal cubic limit, and construct the Hamiltonian for the
case with octahedral rotations by making local unitary rotations on the $j\!=\! 1/2$ spins which
induces Dzyaloshinskii-Moriya interactions, leading to an AFM with `weak' ferromagnetism \cite{Jackeli2009,Senthil2011}.

\begin{figure}[t]
\includegraphics[scale=0.35]{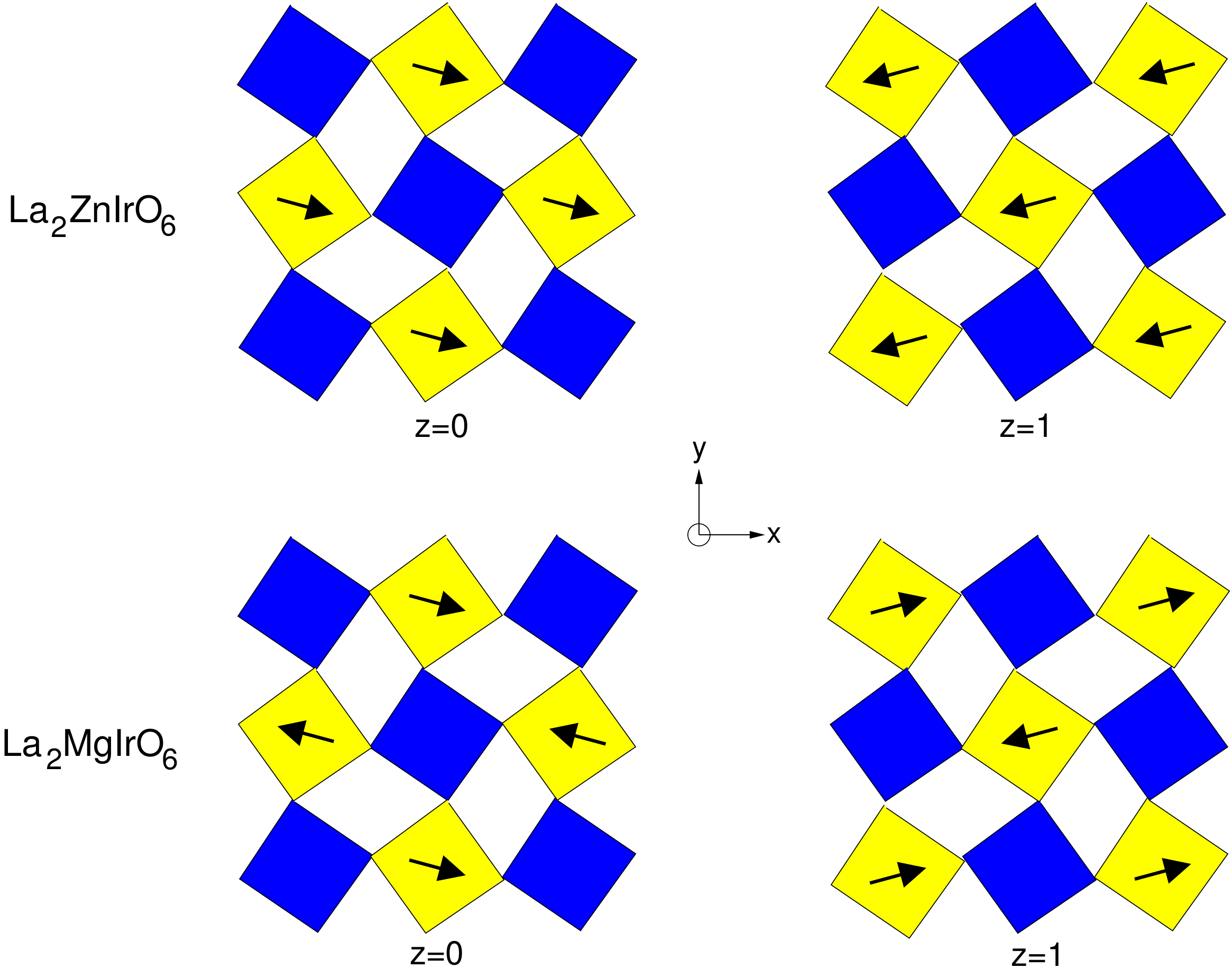}
\caption{Conjectured alignment of staggered octahedral rotations and the ferromagnetic planes in the A-II AFM state for 
La$_2$ZnIrO$_6$ (top) and La$_2$MgIrO$_6$ (bottom), with the spins shown on Ir octahedra (yellow). We have picked the $z$-axis 
as the direction along which the octahedral rotations are staggered for the Ir octahedra, and shown only $z=0,1$ planes. For
La$_2$ZnIrO$_6$, the FM planes are the $xy$-planes stacked antiferromagnetically along $z$, leading to a net `weak' 
ferromagnetic
moment along $-\hat{y}$, while for La$_2$MgIrO$_6$
the FM planes are $xz$-planes stacked antiferromagnetically along $y$ leading to no net ferromagnetic moment. The uniform
Ir octahedral tilts are unimportant for this discussion and is not shown.}
\label{fig:rotn}
\end{figure}

In La$_2$MgIrO$_6$,  {\it ab initio} studies predict a `weak'
ferromagnetic moment $\approx\! 0.3\mu_B$ in the monoclinic P$2_1$/n structure; 
however, experiments do not detect {\it any} ferromagnetic
moment in the ordered phase. To understand this discrepancy we propose that the axis of the 
staggered octahedral rotations and the stacking direction of the ferromagnetic planes are along
orthogonal cubic axes (see Fig.~\ref{fig:rotn}), and {\it ab initio} results may have missed the correct ordering due to subtle energy
differences. This can be tested if additional magnetic Bragg peaks can be resolved using
high resolution X-ray diffraction.
If we ignore SOC ($J_K\!=\!0$, $\Gamma\!=\!0$), and note that $\phi \! \approx \! 9^\circ$  from the
structural data is small, the measured $\Theta_{CW} \! \approx \! -24K$
yields $J_H \! \approx \! 8K$. Our Monte Carlo simulations at $J_K\!=\!0,\Gamma\!=\!0$ show $T_c \! \approx \! 0.44 J_H S^2$,
consistent with previous work on the fcc Heisenberg model \cite{Diep1989}.
Heuristically replacing the classical $S^2$ by $S(S+1)$ for quantum spins leads to a renormalized
$\tilde{T}_c = T_c (1+1/S)$. This is a good approximation
for the 3D cubic lattice $S\!=\!1/2$ Heisenberg model \cite{Sandvik1998}. Here, on the fcc lattice, with $J_H\! =\! 8K$ and 
$S\! =\! 1/2$, we find $\tilde{T}_c \! \approx \! 2.6K$, much smaller than
$T^{\rm expt}_c \! =\! 12K$. With $\Gamma \!\neq \! 0$, but keeping $J_K\!=\! 0$, $T_c$ hardly changes or
even gets suppressed. This 
hints at a significant $J_K \!>\! 0$. Indeed, the ``frustration parameter'' $f= -\Theta_{CW}/\tilde{T}_c$, plotted in Fig.~\ref{fig:Tc}(d) 
for $\Gamma\!=\! 0$, shows that recovering the experimentally
observed small $f \! \approx \! 2$ needs a large Kitaev exchange $J_K/J_H \gg 1$.

Thus, we suggest that a model with a dominant Kitaev term $J_K>0$, perturbed by a weak Heisenberg exchange coupling $J_H \ll J_K$,
is a good starting point to understand $j_{\rm eff}\!=\!1/2$ magnetism in La$_2$MgIrO$_6$; 
we estimate this dominant coupling $J_K \! \approx \! 24K$. These estimates do not shed much light on the off-diagonal
symmetric exchange since the powder averaged $\Theta_{CW}$ is independent of $\Gamma$,
and $T_c$ is not very sensitive to $\Gamma$ (see Fig.~\ref{fig:Tc}(c)). However $|\Gamma| > J_K/2$
is precluded by the observed order. Traditionally, in fcc magnets, robust A-type order
is ascribed to second-neighbor Heisenberg interactions \cite{98_seehra,01_lefmann}.  For heavy oxides, however, our results show that the A-type
AFM, and the small frustration parameter, is due to SOC-induced Kitaev interactions.

In La$_2$ZnIrO$_6$, there is a measured `weak' ferromagnetic moment $\approx 0.22 \mu_B$; thus, the 
axis along which the ferromagnetic planes are stacked in staggered fashion must
coincide with the axis of the staggered octahedral rotations (see Fig.~\ref{fig:rotn}).
Setting $\phi \! \approx \! 11^\circ$, consistent with structural data, we expect a moment $\approx \! 0.19\mu_B$,
close to the measured value. This is smaller than the {\it ab initio} prediction $\approx \! 0.5 \mu_B$. 
Based on the smaller $T_c^{\rm expt} \! \approx\! 7.5K$ in La$_2$ZnIrO$_6$, and assuming similar ratios of
exchanges, $J_H/J_K \! \ll \! 1$, we estimate the dominant 
$J_K \! \approx \! 15K$, and $\Theta_{CW} \! \approx\! -15K$;
however, experiments report $\Theta_{CW} \! \approx \! -3K$ \cite{cao2013}.
This discrepancy remains to be resolved.

In summary, DP Mott insulators are a distinct class of materials which host 
strong Kitaev exchange interactions.
Our study calls for a microscopic understanding of the AFM Kitaev exchange,
motivates a search for DPs with large $\Gamma$, which can stabilize stripes or complex
spiral orders. The A-II AFM order we find in the AFM Kitaev model is stable against quantum
fluctuations for $j=1/2$ moments; a detailed study of quantum fluctuation effects will be reported
elsewhere \cite{Aczel2015b}.

We thank G. Chen, J. P. Clancy, B. D. Gaulin, G. Jackeli, J. E. Greedan, Y. B. Kim,
Y. J. Kim, and S. Trebst for useful discussions. We acknowledge support from NSERC of Canada (AMC,CH,AP),
the Bonn-Cologne Graduate School of Physics and Astronomy (SM), and the Scientific User Facilities Division
of the US Department of Energy, Office of Basic Energy Sciences (AAA).


\end{document}